# Outan: An on-head system for driving μLED arrays implanted in freely moving mice

Alexander Tarnavsky Eitan*, Shirly Someck*, Mario Guillermo Zajac, Eran Socher, Eran Stark

* These authors contributed equally to this work

***Abstract*—In the intact brain, neural activity can be recorded using sensing electrodes and manipulated using light stimulation. Silicon probes with integrated electrodes and μLEDs enable the detection and control of neural activity using a single implanted device. Miniaturized solutions for recordings from small freely moving animals are commercially available, but stimulation is driven by large, stationary current sources. We designed and fabricated a current source chip and integrated it into a headstage PCB that weighs 1.37 g. The proposed system provides 10-bit resolution current control for 32 channels, driving μLEDs with up to 4.6 V and sourcing up to 0.9 mA at a refresh rate of 5 kHz per channel. When calibrated against a μLED probe, the system allows linear control of light output power, up to 10 μW per μLED. To demonstrate the capabilities of the system, synthetic sequences of neural spiking activity were produced by driving multiple μLEDs implanted in the hippocampal CA1 area of a freely moving mouse. The high spatial, temporal, and amplitude resolution of the system provides a rich variety of stimulation patterns. Combined with commercially available sampling headstages, the system provides an easy to use back-end, fully utilizing the bi-directional potential of integrated opto-electronic arrays.**

## I. INTRODUCTION

UNDERSTANDING the function of neuronal circuits in the brain requires an ability to both record and control the activity of neurons. Experiments with freely moving (in contrast to head-fixed) animals are adequate in terms of behavior and brain states [1]–[3] but impose strict constraints on the size and weight of the equipment rigidly connected to the head of the animal [4]–[6]. Extracellular electrophysiological recordings can be used to monitor and detect neuronal activity [7], [8]. Optogenetics is a technique that allows neuroscientists to manipulate neural activity using light in the brain of genetically modified animals [9]. While dense arrays of electrodes for extracellular recordings in freely moving animals have been an established technique for many years, precise light delivery for optogenetics is still a challenge [10], [11]. Recent advances in micro electromechanical systems (MEMS) fabrication allowed miniaturization of light sources to dimensions suitable for injection or implantation (μLEDs) [12], [13]. Furthermore, silicon probes integrating 10 x 15 μm μLEDs with a dense electrode array have been developed [14]–[16] and used in freely moving animals [14], [17]. **Fig. 1A** shows a commercially-available probe designed for combining optogenetics with extracellular recordings (NeuroLight Technologies, USA), packaged as an implantation-ready device.

In addition to the implantable device, an electronic "back-end" system is required to perform recording and stimulation. For recording, the commercially-available RHD2000 series of headstages (Intan Technologies, USA) are widely used with various implanted devices [18]. In contrast, for optogenetic stimulation, miniaturized current drivers have been reported [19]–[27], but to the best of our knowledge, none of these miniaturized current drivers have been used for combining high-density electrophysiological recordings with multi-site optical control in freely-moving animals. It is difficult to drive integrated μLEDs for two reasons. First, the small area of the blue μLEDs results in a high forward voltage (**Fig. 1D**), requiring the driver to output 5 V to achieve a current of 1 mA. Second, in probes combining μLEDs with dense electrode arrays ([14]–[16]), the cathodes of all μLEDs are connected to a single ground net (**Fig. 1C**), requiring the driver to source rather than sink the currents. Previous reports have demonstrated miniaturized drivers that support only larger μLEDs or provide insufficient output power. A stimulation platform for freely moving rodents is described in [19] but the stimulation is performed by 1 mm LEDs. The circuits presented in [20] and [21]–[23] are designed for 200 μm μLEDs with a forward voltage of 3 V @ 20 mA and sink the current. The devices proposed in [24]–[26] combine a driver with a probe on the same IC with options for 20 μm and 200 μm μLEDs, but animal tests were performed only with 200 μm μLEDs. The μLED driver in [27] targets 10 x 15 μm μLEDs and was designed to source 1 mA, but measurements were demonstrated only up to 100 μA, supporting only a tenth of the dynamic range of the μLED light power. In addition, the IC was bonded rigidly to the probe, limiting its use to anesthetized and not freely moving animals. At present, all neuroscience research published using the NeuroLight probe (and other integrated 10 x 15 μm μLED probes) used large stationary drivers connected to the preparation through long overhanging cables [14]–[17], [28]. Due to the exponential nature of the I-V profile of the μLEDs and the small currents required, small variations in the

---

Manuscript received December 10, 2020. This work was supported by an ERC grant (#679253) and a Rosetrees grant (#A1576). Alexander Tarnavsky-Eitan, Shirly Someck, and Eran Stark are with the Sagol School of Neuroscience and with the Department of Physiology and Pharmacology, Sackler Medical School, Tel Aviv University, Tel Aviv 6997801, Israel. Mario

Guillermo Zajac and Eran Socher are with the Department of Physical Electronics, The Iby and Aladar Fleischman Faculty of Engineering, Tel Aviv University, Tel Aviv 6997801, Israel. Correspondence: eranstark@tauex.tau.ac.il.

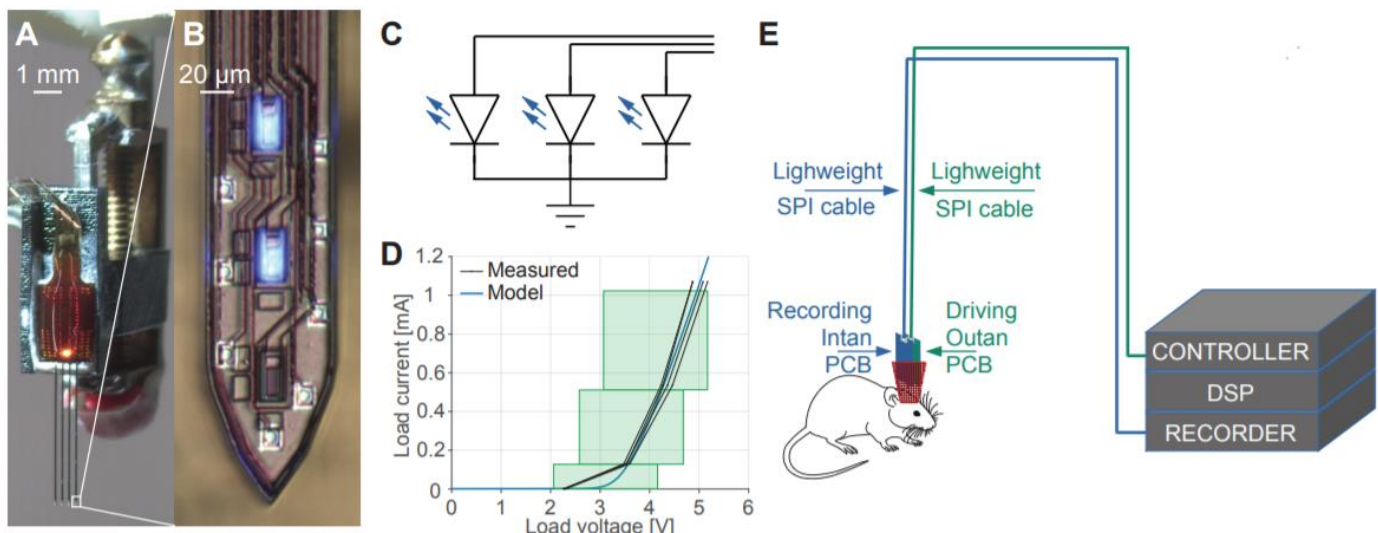

**Fig 1. Design goal and system conceptualization. (A)** Implantation-ready µLED opto-electronic probe, mounted on a micro-drive. **(B)** Tip of a single shank of a four-shank silicon probe with integrated µLEDs (NeuroLight Technologies). Each shank consists of 8 recording sites and 3 µLEDs; the top two µLEDs are on. **(C)** Electrical schematic of the µLEDs in a single shank. **(D)** I-V profile of a µLED. The black lines connect SMU measurements of four different µLEDs, and the blue line is a fit to a model consisting of a diode in series with a resistor. The green rectangles are regions of operation of the ASIC current source (**Figure 2**). **(E)** Schematic representation of the experimental setup.

anode voltage accumulated on the overhanging cable cause large errors in stimulation intensity. The drivers are either custom built current sources or off the shelve voltage generators which cannot control all channels simultaneously.

To create an integrated "back-end" system for the integrated µLED probe (NeuroLight Technologies, USA), we propose miniaturizing the µLED driver into a modular headstage, complementary to the Intan headstage used for recording (**Fig. 1E**). An Intan PCB amplifies, multiplexes, and samples (ADC) the signals from the electrodes to the recorder. The "Outan" PCB de-multiplexes controller commands and sources (DAC) current to the µLEDs. The proximity of the current source to the probe prevents EMI and other noise from accumulating on the overhanging cables. Serial digital control of the driver allows driving a large number of independent channels, while maintaining a wide dynamic range and high resolution. The long overhanging cables carry power and low-voltage differential signaling (LVDS) of serial peripheral interface (SPI) protocol to the headstage. The control signals can be transmitted using lightweight cables because digital communication is much more robust to noise.

Section II describes the design of an application specific integrated circuit (ASIC) current source designed to drive 32 µLEDs with up to 1 mA per channel. Section III describes the integration of the ASIC with required peripheries into a modular PCB headstage we call "Outan". The headstage is roughly the same size as the Intan headstage. Section IV includes measurements of the performance of the system against µLED probe loads. To demonstrate functionality, the headstage was used with a freely moving mouse. These experiments are described in Section V. The system succeeded in activating individual neurons as well as inducing oscillations in local populations. Control over multiple µLEDs allowed generating synthetic multi-neuronal patterns at arbitrary sequences. As discussed in Section VI, the Outan headstage in combination with the Intan headstage provide a complete back-end suite for full utilization of the research potential of the integrated silicon µLED probes.

## II. CURRENT SOURCE ASIC

We designed a current source device to support the operation of µLED opto-electronic probes in freely-moving mice (**Fig. 1**). The design specifications for the device are listed in **Table I**. To make the device as small and power-efficient as possible, an ASIC was designed and manufactured in TSMC 65 nm process. The architecture of the ASIC is described in **Fig. 2A**. The chip consists of a low-voltage band-gap current reference based on [29] and 32 channels which are identical to each other, except for a hardcoded address. The reference module also includes a series of high-swing cascode current mirrors to generate all required bias voltages. To ensure good matching, the current mirrors in the generator have the same parameters as the cascodes they bias. A detailed description of the common generator for reference and bias voltages is available in [30].

The typical use case for the system is sending short activation events (e.g. pulses or sinusoids) to the µLEDs. In such a case, most µLEDs are turned off most of the time and in a single probe scenario, only 12 out of the 32 channels are connected to µLEDs. To take advantage of the sparsity of activation, a random access control scheme was chosen. Each command includes the number of the accessed channel (a 5 bit address) and the output value required of the channel (encoded using 10 bits). To comply with standard SPI controllers, an additional don't-care bit was added for a total of 16 bits. The output of a channel is kept constant in a zero-order-hold manner as long as the command is not addressed to that specific channel. The specification of a 5 kHz refresh rate per channel limits the zero-order-hold to 0.2 ms, minimizing any potential thermal dissipation by the load.

TABLE I
ASIC DESIGN SPECIFICATIONS

| Specification | Value |
|---|---|
| Driver type | Current source (not sink – **Fig. 1C**) |
| Designated load | 10 x 15 µm µLED (5 V @ 1 mA) |
| Number of channels | 32 |
| Maximal output per channel | 1 mA |
| Resolution | 10 bits |
| Refresh rate per channel | 5 kHz |
| Maximal total output | 32 mA |

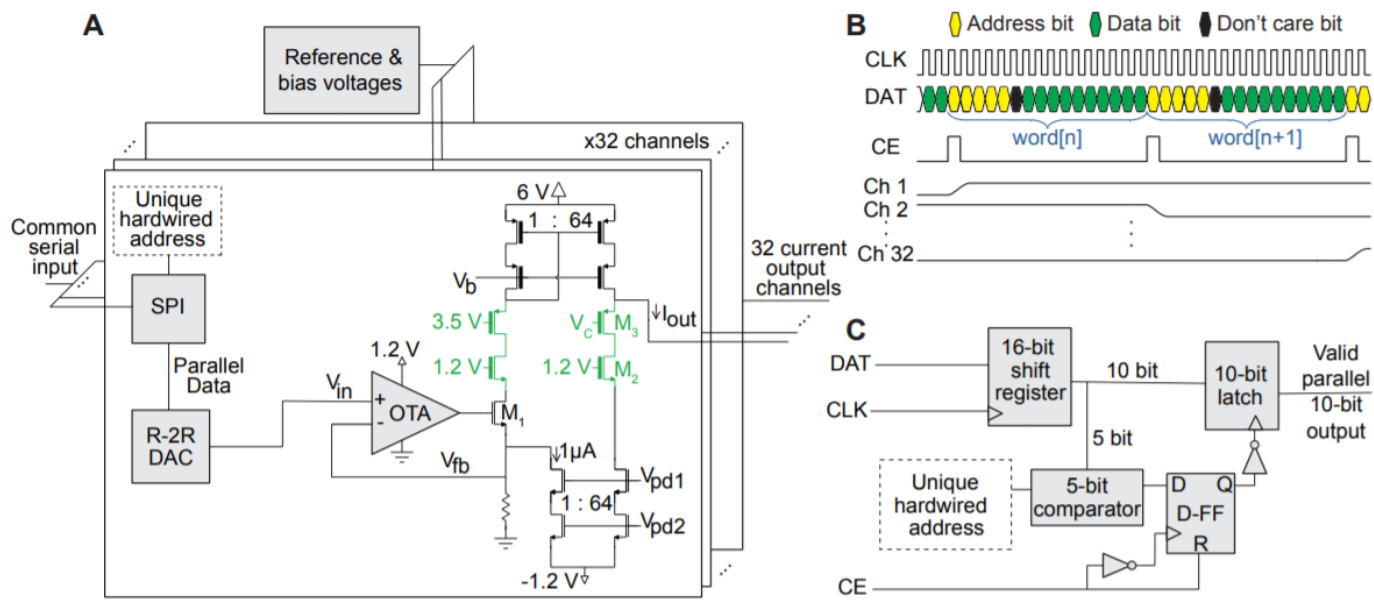

**Fig 2. Design of the current source ASIC.** (**A**) Schematic of the ASIC. High Voltage (HV) transistors have thick gate lines. Voltage clamps marked in green. (**B**) Modified SPI communication scheme using a continuous clock. (**C**) Block diagram of the modified SPI module.

**Fig. 2A** describes the chosen architecture for the current control. Due to the proximity of the small µLEDs to the recording sites (15 µm), voltage changes on the anodes of the µLEDs during the onset and offset of stimulation cause artifacts on the recording channels. This makes PWM control of the µLEDs prohibitively noisy, requiring analog current control.

Each current source channel receives all the commands, and the channel-specific SPI module responds only if the channel is addressed by outputting the required current value in 10 parallel 1.2 V digital signals. These parallel data are converted into a single analog voltage between 0 and 0.38 V by a 10 bit non-segmented R-2R digital to analog converter (DAC). Then, the analog voltage is converted to analog current by an operational transconductance amplifier (OTA) with an NMOS transistor and a feedback resistor. Based on experience gained from preliminary designs and for simplicity, an R-2R with OTA was chosen over a current steering architecture. The feedback resistor makes the gain dependent on temperature. To cancel this dependency, the R-2R is based on a voltage reference that depends on temperature in an opposite manner. To support load compliance of up to 5 V, the current from transistor $M_1$ is multiplied by a 1:64 current mirror, which also lifts the current to a high (up to 6 V) power supply. Note that in practice, load compliance is limited by the high power supply provided to the chip. For the Outan headstage PCB the high supply is 5 V, limiting system compliance to 4.6 V (see Section III).

The SPI module in **Fig. 2A** implements a modified version of the standard SPI protocol. Digital commands are sent to the system in a feed-forward scheme described in **Fig. 2B**, without requiring a master-in-slave-out (MISO) line. In this case, in contrast to standard SPI, the clock and data signals can operate continuously without wasting a clock cycle while CE is high. Eliminating the wasted clock cycle allows a 6% (1/17) shorter command time. A block diagram of the implementation is shown in **Fig. 2C**. The latch outputting the parallel data to the DAC is triggered only by the rise of CE and is not affected by the fall of CE, allowing the SPI block to work with the standard SPI as well as with the modification described above.

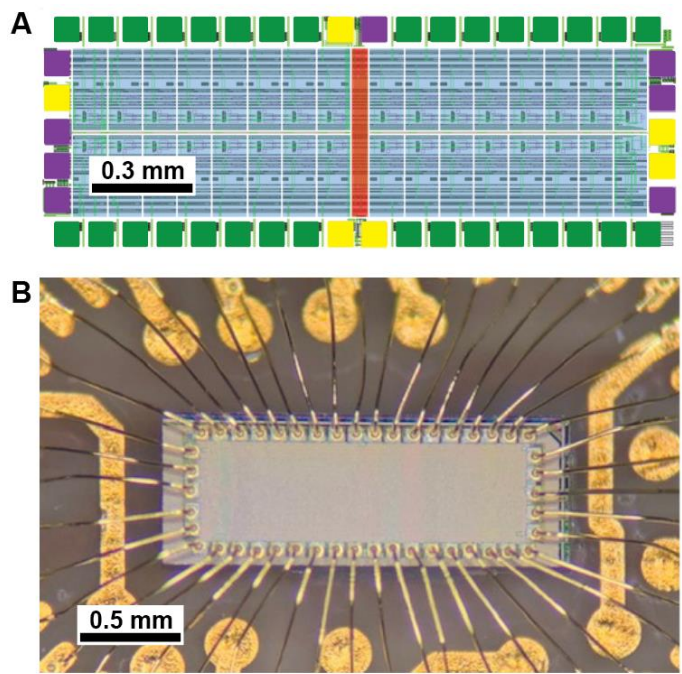

**Fig 3. ASIC implementation.** (**A**) Layout of the top metal layers of the ASIC. The 32 channels are tinted blue; the common generator for reference and bias voltages are tinted red; the current output pads are tinted green; the analog signal pads are tinted purple; the pads for digital signals and supplies are tinted yellow. (**B**) Photo of the fabricated and wire-bonded ASIC. The size of the chip including pads is 2 by 0.7 mm.

The feedback transistor in the output stage ($M_1$ in **Fig. 2A**) would cut off if the current through it is too small, opening the feedback loop for small outputs. For the minimal non-zero command, the output of the R-2R ($V_{in}$) is 0.38 V / 1024 = 0.37 mV, the required output is 1 µA, and the current through $M_1$ needs to be 1 µA / 64 = 15.6 nA. To prevent the cutoff, a constant bias current of 1 µA is added to the input side of the current mirror, compensated by subtracting 64 µA from the output side of the current mirror. This bias current keeps $M_1$ in saturation but requires the usage of a negative supply. The subtraction current sources must be linear to correctly subtract the multiplied biasing current from both sides of the mirror. A single transistor source does not have high enough output impedance, so cascode sources are used. To separate the negative voltage domain, a clamp transistor was added in the bias generator. The resulting three transistor stack requires a negative supply of -1.2V. The current subtracting transistors and their cascode bias generators ($V_{pd1}$, $V_{pd2}$) are implemented in a deep n-well as the substrate is grounded.

All the transistors in the SPI module, R-2R DAC and the OTA are described by the TSMC process documentation as "core" transistors and are rated up to 1.2 V. For the output stage, we used special high voltage (HV) transistors which are rated up to a $V_{DS}$ of 5 V and $V_{GS}$ of 2.5 V. The combination of up to 6 V supply required by the 5 V forward voltage of the load and the -1.2 V supply for the subtraction sources creates a total voltage drop of 7.2 V on the output stage. This drop is too high even for HV transistors. Therefore, voltage clamps were added (green transistors in **Fig. 2A**) to protect the core transistors and the HV transistors from the maximal possible voltage drop of 7.2 V. The gates of the clamps are biased to a constant voltage, putting a lower bound on voltages of the source of PMOS clamps, and an upper bound on voltages of the source of NMOS clamps.

The gate voltage of the output clamp cannot be set to a constant value. For zero output current, the output voltage must be below 2.5 V, demanding a gate voltage below 2.5 V for the clamp. However, when outputting the maximum current, the output voltage must be 5 V, requiring the same gate voltage to be above 2.5 V (to comply with the $V_{GS}$<2.5 V rating). To resolve the contradicting demands, instead of a constant gate voltage, the gate of the output clamp is set to one of three values selected by a current steering DAC depending on the received digital command. This solution limits the system to work only against loads which fit into the green rectangles in **Fig. 1D**.

In the physical layout of the ASIC (**Fig. 3A**), the 32 channels are arranged in two rows to minimize distance to output pads. A common generator for reference and bias voltages is positioned in the middle. Analog supply pads are duplicated to ensure even distribution. The digital elements were implemented inside a deep-n-well, powered by a separate "digital" supply, and surrounded by a separate "digital" ground disconnected from the ground of the analog circuits to prevent high frequency noise. ESD protection diodes were added to the input pads to allow safe handling of the system. The resulting layout floor-plan is described in **Fig. 3A**, showing the 32 channels, reference generator and surrounding pads. The total active area for each channel is 250 x 100 µm, with the largest part being the 1:64 multiplying mirror at 70 x 100 µm. The digital elements in each channel are concentrated in a 50 x 50 µm square. The DAC resistors are implemented as a tight array of identical segments with a total area of 30 x 35 µm. The fabricated chip has dimensions of 2 x 0.7 mm including bonding pads (**Fig. 3B**). The idle power consumption of the ASIC in simulation is 16 mW.

## III. INTEGRATION INTO A HEADSTAGE

The ASIC is too small to be handled manually and requires some electrical peripherals in order to be connected to the controller. To accommodate both requirements, a custom PCB was designed, serving as a headstage for the ASIC. Initially, a breakout chip with only two channels and 28 test pads for measuring internal signals was fabricated to test the circuit and the electrical peripheries required by the ASIC. Testing the breakout chip in a probe station revealed that if the high supply is turned on before the analog and digital supplies, internal voltages stop responding to digital commands, and the chip does not source any current. This result was permanent, indicating that irreversible damage to the chip occurred. To prevent damage to the chip, the PCB was designed with peripherals, controlling the power-on and power-off sequences of the supplies. The analog supply (1.2 V) is powered on first, then the digital supply (1.2 V), and finally the negative (-1.2 V) and high (5 V) supplies together. The power-off sequence is reversed relative to power-on. Another issue for the periphery to solve is signal translation. To protect the signal from interference, the long SPI wires in **Fig. 1E** conduct signals encoded using low voltage differential signaling (LVDS). In contrast, the logic transistors in the ASIC operate using 1.2 V logic levels, so a conversion between the two signaling schemes is required.

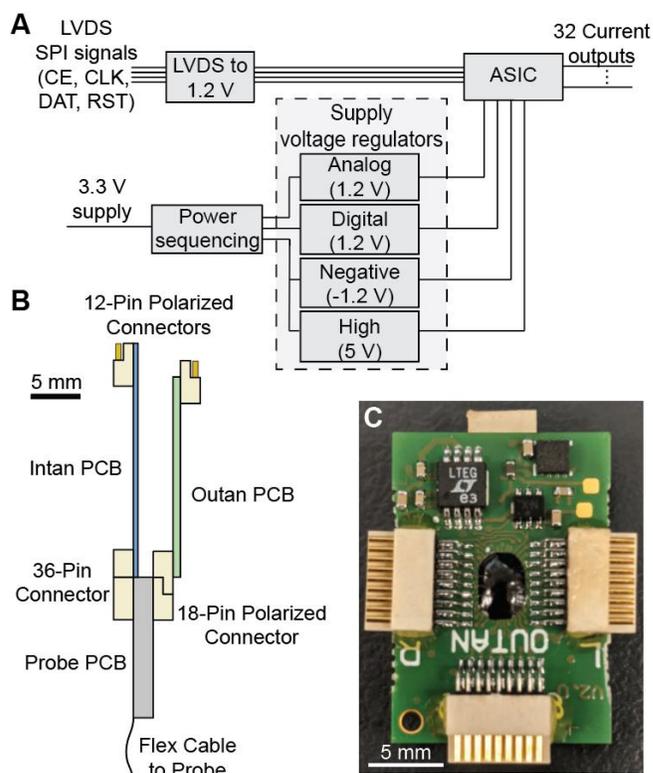

**Fig 4. PCB electrical and mechanical schematic. (A)** Block diagram of the PCB. **(B)** A mechanical schematic of our Outan PCB and an Intan RHD2132 PCB (Intan Technologies, USA) connected simultaneously to an integrated probe (NeuroLight Technologies, USA). **(C)** Photo of the PCB. The assembled PCB is 14 x 20.2 mm, and weighs 1.37 g.

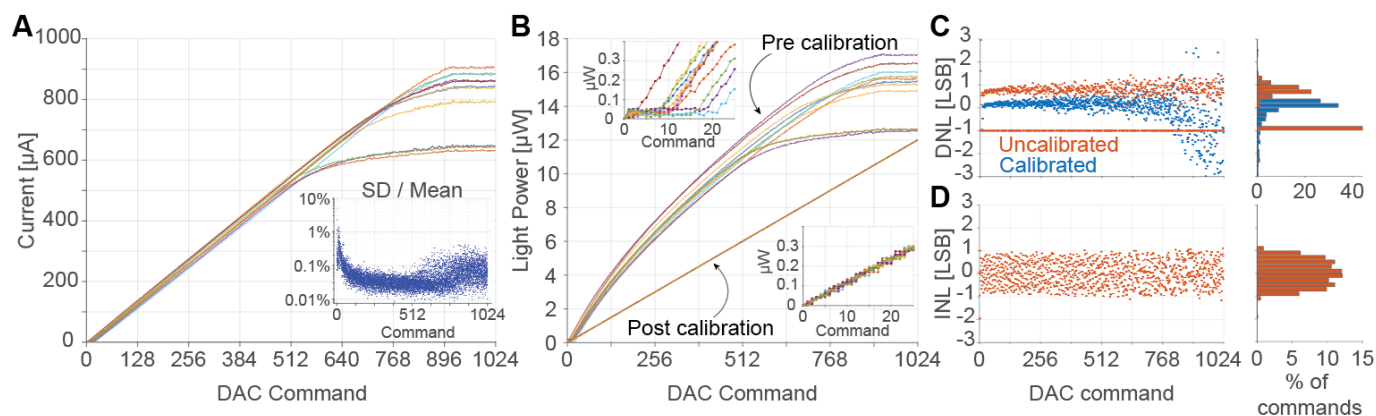

**Fig 5. System allows linear performance.** (**A**) Measured current output for 11 channels (one of the 12 μLEDs failed). Inset shows SD/mean of repeated measurements of the current for each command is < 1%. (**B**) Measured light power output for the same channels as in **A**. Digital calibration fixes the compliance-limited non-linearity and the variance between channels. Insets show the lower end of the scale before (top) and after (bottom) calibration. (**C**) Differential non-linearity (DNL) for one channel, calibrated to 0.6 mA maximum output. (**D**) Integral nonlinearity (INL) for the same channel as in **C**.

From a mechanical perspective, the Outan PCB has to provide connectors and be compact enough to fit in the headstage assembly. In addition to the Outan PCB, the headstage assembly includes two commercially-available PCBs. First, the commercially available silicon probe (**Fig. 1A**; NeuroLight Technologies, USA) arrives connected by a flexible cable to a PCB equipped with two connectors (Omnetics Connector Corporation, USA). The 12 μLEDs are routed to one connector (18 pin polarized), and the 32 recording sites are routed to another connector (36 pin). Second, for recording, an Intan RHD2132 PCB (Intan Technologies, USA) is attached vertically to the probe PCB. The Outan PCB must be connected to the same probe PCB together with the Intan PCB, while preventing mechanical collisions and allowing easy access to both cable connectors.

The block diagram of the Outan PCB is shown in **Fig. 4A**. The Outan PCB includes four voltage regulators which are turned on and off in the order of power-up and power-down required by the ASIC. The control of the regulators is performed by switching their enable signals using a sequencing IC. In case of a sudden disconnection, a large (200 μF) capacitor holds the input voltage high long enough for the power-down sequence to finish. Transistor switches were added to the output of each regulator to quickly ground the voltage during shut down. The Outan PCB also includes an LVDS receiver that outputs 3.3 V signals and a translator from 3.3 V to 1.2 V signaling. In order to simplify the design, instead of the required 6 V supply, a 5 V supply is provided to the chip, limiting the output voltage to 4.6 V and preventing the PCB output current from reaching the full 1 mA supported by the ASIC. A load compliance of 4.6 V is unproblematic for the target application, since over 10 μW were outputted on every single μLED channel – much higher than required for generating both single-unit and local population activity in the brain (see Section V).

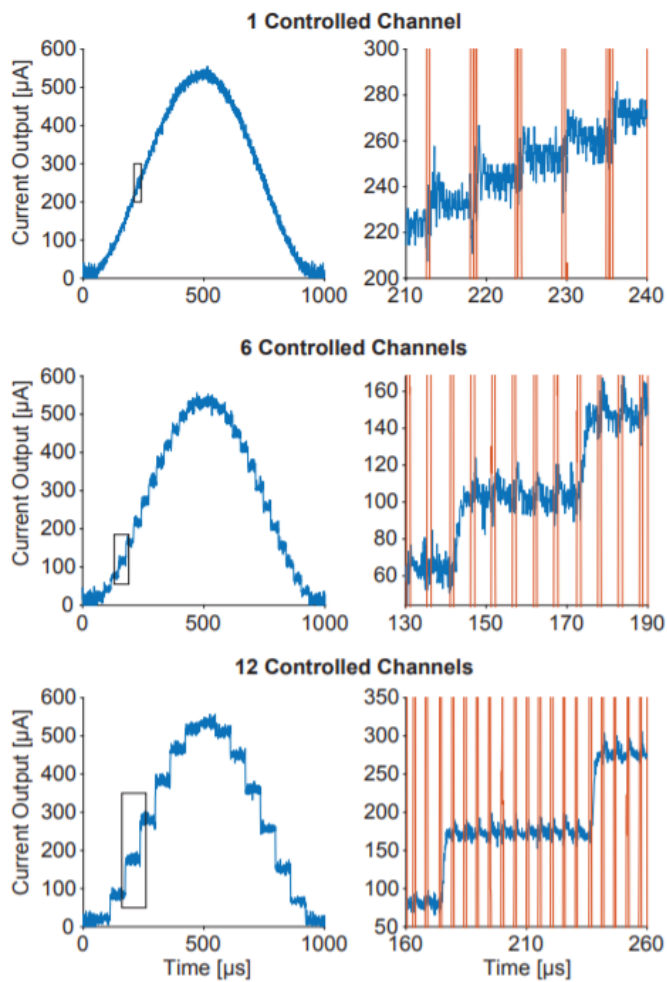

**Fig 6. System allows sub-millisecond control of multiple channels.** Measured output current of a single channel for 1 kHz sinusoid commands while controlling a total of 1, 6 or 12 channels. On the right are enlarged views of the steps in each case. Orange vertical lines depict the times of SPI command communication. There are six commands between changes to the measured channel when six channels are controlled in parallel (middle panel), one for the top, and 12 for the bottom panels.

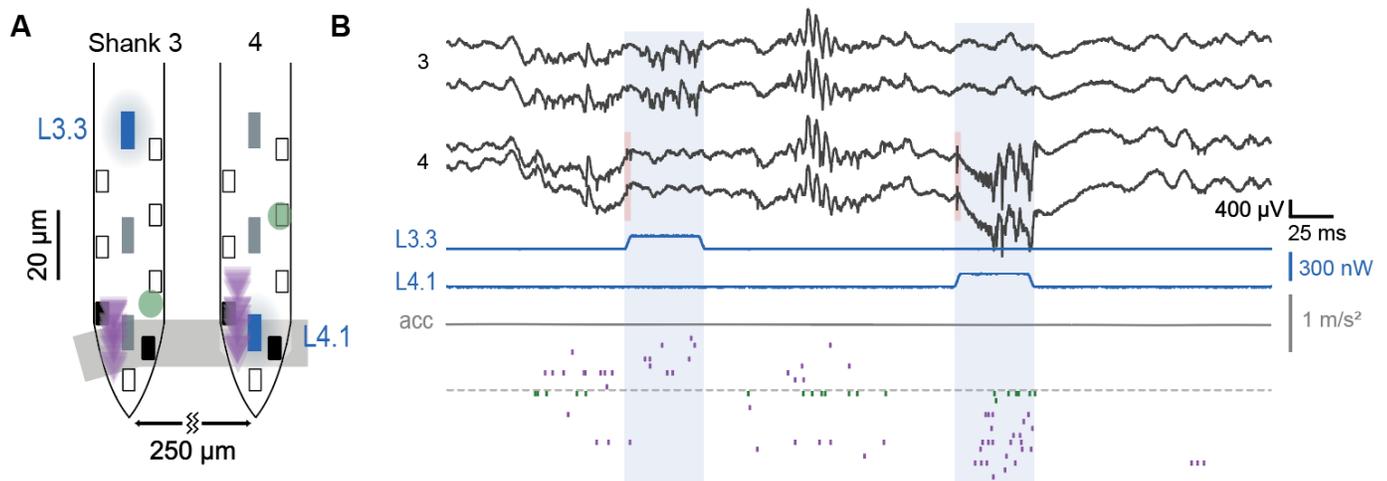

**Fig 7. System allows wide-band recordings and multi-channel stimulation in freely-moving mice.** (**A**) Schematic of the µLED probe that was implanted in the CA1 pyramidal layer (indicated by the wide grey trace), recording multiple putative pyramidal cells (PYR; purple triangles) and parvalbumin-immunoreactive cells (PV; green circles). The probe consists of two shanks, and each shank has eight neural recording channels. Two of the channels are marked in black, and only the signals recorded from these channels are presented in panel **B**. (**B**) Wideband (0.1–7,500 Hz) neuronal activity monitored by four sites (marked in black in panel **A**) during sequential low-power illumination (peak power, 150 nW) of two µLEDs (marked in blue in **A**). Onset artifacts are highlighted in pink. Spikes of PYR (purple) and PV (green) are indicated by the tick marks in the raster plots below the acceleration trace.

A mechanical diagram of the connections to the probe PCB is shown in **Fig. 4B**. The Intan PCB has a 36 pin connector on the bottom end for the electrodes and a 12 pin polarized connector on the top end for an LVDS SPI cable. Similarly, the Outan headstage PCB connects to the 18 pin polarized connector on the bottom side and has a 12 pin polarized connector for an LVDS SPI cable. The Outan PCB was designed to have a length similar to the Intan PCB so that the connectors of each can be accessed separately. The fragile implanted silicon probe is mechanically separated from the PCBs by a flexible cable. The flex cable allows fixating the probe PCB rigidly to the skull and mounting the probe itself on a movable micro-drive (**Fig. 1A**). This allows connecting external cables to both SPI connectors, as required for using the system in freely moving animals. The connectors on the Outan PCB are placed on opposite faces of the board to allow both headstage PCBs to be connected to the probe PCB without collisions. The Outan PCB has two additional 18 pin polarized connectors on the sides, which enables driving two additional probes with a single PCB (36 µLEDs). The ASIC has 32 channels, so four of the channels are wired to two µLEDs each, one on the left connector and one on the right. Driving multiple probes requires connection jumpers because the additional probes will be implanted in different locations in the brain, so side connections are most comfortable.

As can be seen in the photo of the Outan PCB (**Fig. 4C**), the ASIC is bonded directly to the Outan PCB. A 1.2 mm through-hole ground via is located in the corner of the PCB for possible mechanical fixation and electrical grounding. Test pads are exposed for all the inputs to the ASIC - the outputs of the voltage regulators and the four digital 1.2 V signals. The size of the Outan PCB is 14 x 20.2 mm, and the weight is 1.37 g. The idle power consumption of the whole system (including voltage regulators) was measured at 84 mW, with additional 6.95 mW/mA during stimulation.

## IV. Bench Experiments Demonstrating Linear Control of Multiple Channels

### A. Linear control of light power with variable load properties

The variance in fabrication of the µLEDs results in different forward voltages and light profiles for each individual µLED, and some may even fail completely. Linear control of the load current would result in nonlinear light power output for a single µLED, and in distinct light power levels for different µLEDs when driven with identical currents. Calibration of the light output for each µLED helps achieve linear and uniform control of the output light power.

To implement linear calibration, the current and light power outputs of the Outan PCB were first measured with an implantation-ready silicon probe as the load. The mean output current for each command on every channel are presented in **Fig. 5A**. Since the Outan PCB limits load compliance to 4.6 V, the output current saturates before 1 mA. For different µLEDs, saturation occurs at different values due to inter-µLED variability of the I-V profiles. For some µLEDs, 0.6 mA corresponds to 4.6 V; whereas for other µLEDs, 0.9 mA corresponds to 4.6 V. Since a code of 1024 corresponds to 1 mA, for µLEDs of first type, saturation begins around code 512, whereas for the second, saturation sets in later (around code 896). Correspondingly, the light power output (**Fig. 5B**) varies between µLEDs, ranging from 12.5 to 17 µW. In addition to saturation at high values, light output is not linear at the low end of the scale: power stays constant for the first 5-20 commands and then starts to rise (see **Fig. 5B**, top left inset).

For each channel, the standard deviation (SD) of the current was calculated across 9 measurements with a 6.5 digit multimeter for each DAC command. For most (88%) commands, the SD is less than 0.1% of the mean. The coefficients of variation (SD/mean) of repeated measurements of the current are presented in the inset of **Fig. 5A**.

Digital calibration can compensate for intra-µLED

nonlinearities, inter-µLED variance, and variability between the ASIC channels, creating a linear and uniform output profile for all µLEDs (**Fig. 5B**, post-calibration; see also **Fig. 5B**, bottom right inset). The calibration is performed at the controller level, and no changes to the circuitry are made. The calibration table stored in the controller is 1024 rows deep (encoding desired output in 10 bits) by 32 columns wide (one per channel). First, a range of desired output values in µW is assigned to the 1024 rows in the table. For every combination of channel and desired output (in µW), the entry in the table contains the command for which the closest output was previously measured with a photometer from that channel. Thus, the table is constructed for a specific combination of PCB and probe, addressing all possible end-to-end nonlinearities. Notably, calibration can only prevent but cannot solve saturation. To obtain a linear curve without saturation, we calibrate all channels to the maximum output of the weakest channel (in **Fig. 5B**, command 1023 encodes 12 µW). In that manner, no channel is required to deliver an output above its own maximum, and saturation is prevented.

Fig. 5C shows the differential nonlinearity (DNL) for a representative channel before and after calibration and a histogram of the DNL values. The DNL pre-calibration gets >1 LSB upon reaching the compliance point. Calibration splits the DNL, achieving a lower slope. **Fig. 5D** shows the integral nonlinearity (INL) between the desired profile and the measured results. The absolute value of the INL is larger than 1 LSB in rare cases where no measured value was closer than 1 LSB to the desired value. Three probes were calibrated to 12, 13 and 16 µW maximal outputs, corresponding to LSB values of 11.7, 12.7 and 15.6 nW, respectively.

### B. Random access control of multiple channels

The ASIC was successfully tested with up to 160k commands per second (6.25 µs per command) and may be capable of higher communication rates. A communication rate of 160k commands per second allows simultaneous control of 32 channels at 5k commands per second per channel, consistent with the design specifications (**Table 1**). **Fig. 6** shows a 1 kHz sinusoid measured on one channel, while controlling either a single channel, six, or 12 channels. Measurements were performed with an instrumentation amplifier and a 1 GHz oscilloscope, allowing high temporal resolution at the cost of additional noise. The refresh rate in the typical case of 12 channels is 13.3 kHz, shown as 14 steps during one cycle of the sinusoid (**Fig. 6**, bottom-right).

The cross-talk between channels was smaller than -52 dB (1:417). The worst case scenario cross-talk was between channels adjacent in the probe but not adjacent in the chip. Thus, while not measured directly, the cross-talk between channels of the ASIC may be even lower.

## V. THE SYSTEM ALLOWS GENERATING SYNTHETIC FIRING SEQUENCES IN FREELY MOVING MICE

### A. Surgical and electrophysiological methods

A four-shank integrated µLED probe (NLT-N1-010,

NeuroLight Technologies, USA) was mounted on a microdrive and cleaned in 2.5% Contrad 70 (Decon Laboratories; 60 min in 60°C). Impedance, measured in 0.9% NaCl at 1 kHz (NanoZ, White Matter), was $1.00 \pm 0.28$ MΩ (mean±SD over a total of 32 recording sites). Voltage and power measurements were taken for each µLED at currents from 1-300 µA (2401 source meter unit [SMU], Keithley; PM100D power meter, Thorlabs). A second measurement of the I-P curve was done using Outan instead of the SMU, yielding similar results. The second set of measurements was used to calibrate the µLEDs (**Fig. 5B**).

The complete system was validated in a freely-moving mouse. All animal handling procedures were in accordance with Directive 2010/63/EU of the European Parliament, complied with Israeli Animal Welfare Law (1994), and approved by the Tel Aviv University Institutional Animal Care and Use Committee (IACUC #01-20-049). A 21-week-old male mouse expressing ChR2 in pyramidal cells under the CaMKII promoter was generated by crossing a CaMKII-Cre male (#005359 Jackson labs) with an Ai32 female (#012569 Jackson labs). The mouse was equipped with a 3-axis accelerometer for monitoring head-movements. The probe was implanted in the neocortex above the hippocampus (PA/LM/DV, 1.6/1.1/0.59 mm) under isoflurane (1%) anesthesia following previously-described procedures [31]. The assembly of the headstage was performed during the implantation surgery. The probe, mounted on a microdrive, was inserted to a depth of 590 μm into a craniotomy, and a silicon mixture (3-4680, Dow Corning) was applied to cover the brain and prevent infection. A copper mesh cage was built around the probe for mechanical and electromagnetic shielding, and the recording and stimulating connectors were cemented to the inside of the mesh wall. Both the Intan and Outan PCBs were connected to the probe only during experimental sessions. After recovery from anesthesia, the probe was translated vertically using the screw on the microdrive to the CA1 pyramidal layer over several weeks.

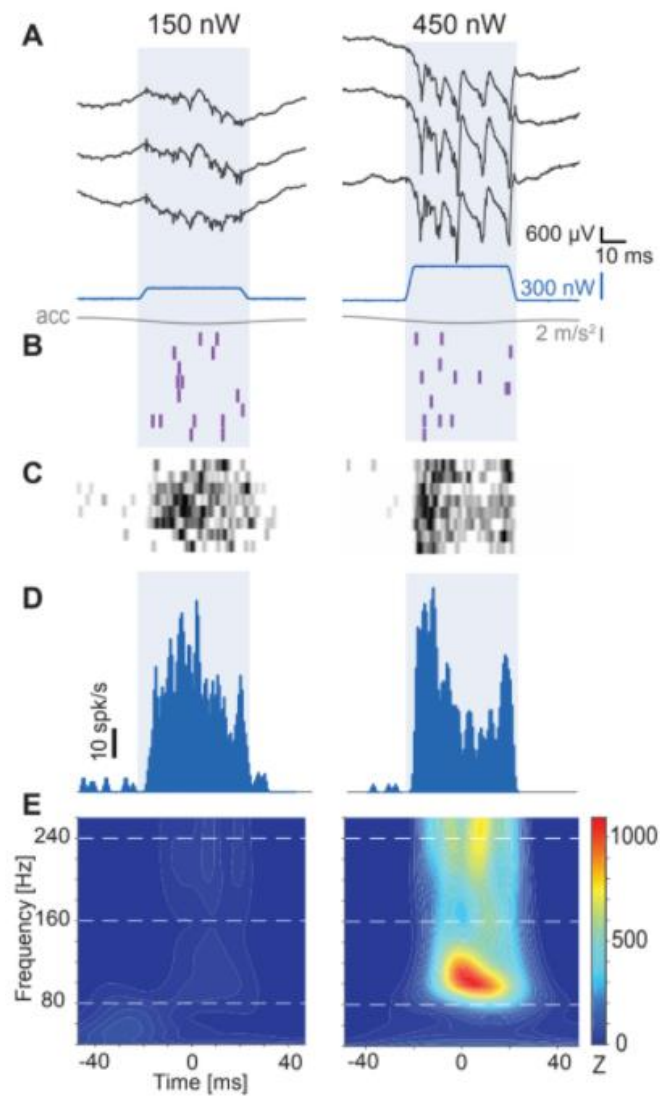

**Fig 8. High resolution and large dynamic range allow both single-unit and population control. (A)** *Left*: Wideband traces recorded from the three bottom-most sites of shank 3 during low-power illumination (150 nW) of the bottom-most µLED (same session as in **Fig. 7**). *Right*: Recording from the same sites during higher-power illumination (450 nW). **(B)** Raster plot with spikes indicated by the tick marks. **(C)** Peri-stimulus time histograms (PSTHs) of eight simultaneously-recorded units (25 stimulation events). For presentation purposes, each histogram was scaled and color coded (0, white; 1, black). **(D)** Mean PSTH, averaged over units. **(E)** Time-frequency wavelet decomposition of the current source density (CSD) recorded from the central site, averaged over the same stimulation events. Power in each frequency band was Z-scored according to the mean and SD power in the lack of illumination.

Recordings sessions were carried out in the home cage during spontaneous behavior, without any specific behavioral task. Neural activity was filtered, amplified, multiplexed, and digitized on the headstage (0.1–7,500 Hz, x192; 16 bit, 20 kHz; RHD2132, Intan Technologies). If spontaneous spiking activity was observed, a full recording session (>3 hours) was conducted, comprised of a baseline period of at least 15 min, followed by response mapping and sequence induction. During response mapping, 25-50 ms light pulses were generated by each diode separately, and the minimal intensity that evoked an observable effect was noted. For data analysis, waveforms were linearly detrended, projected onto a common basis obtained by principal component analysis of the data, and sorted automatically followed by manual curation and re-clustering of noisy units. Only well-isolated units were used for analyses.

### B. In-vivo experiments

First, we tested whether activation of different μLEDs results in distinct neuronal activity patterns. **Fig. 7** shows an example of such activation. Illumination via different μLEDs resulted in high-frequency oscillations (HFOs [14], [32]) in the local field potential (LFP) induced at distinct brain sites. Between the two light pulses occurred a spontaneous high-frequency "ripple" oscillation (HFO), characteristic of CA1 during immobility (head acceleration shown in grey). While units on both shanks spiked during the spontaneous HFO, the low-power illumination pulses affected only local (same shank) firing rates and induced localized HFOs.

The rising and falling edges of the stimulation pulse require a voltage swing on the order of 1 V on the anode of the μLEDs. The sharp change in anode voltage may interfere with the recording electrodes, creating a stimulation artifact [14]. Artifact minimization may be achieved using stimulus waveforms that minimize voltage changes [15]. Flanking 30 ms pulses by two halves of a single cycle of a 100 Hz sinusoid (5 ms on each side) completely prevented both onset/offset artifacts and DC shift from occurring in some channels (**Fig. 7B**, L3.3 illumination) but not in others (**Fig. 7B**, artifacts marked in pink during onset of L4.1 illumination).

Second, we tested whether the resolution of single-channel control is sufficiently high to yield distinct neuronal activity patterns in response to pulses of different light power. In our experiments, the minimal intensity that induced observable spiking was on the order of 100 nW. This is consistent with the proximity of the μLEDs to the cells recorded by the electrodes, and with the dense layout of cells in the pyramidal layer in CA1. Driving the same μLEDs at stronger power levels generated induced HFOs (iHFOs), facilitated by the recruitment of post-synaptic inhibitory neurons [32]. **Fig. 8** shows an example with low power stimulation of 150 nW (just above the spiking threshold), yielding spiking of multiple units but no iHFOs. Applying higher power illumination via the same µLED (450 nW; above the threshold for inducing population effects) induced spiking at a higher rate, organized in oscillatory cycles, and was accompanied by iHFOs (**Fig. 8E**). The full dynamic range of the system, spanning the range of tens of nW to over 10 µW, can be used for other less sensitive opsins and brain regions.

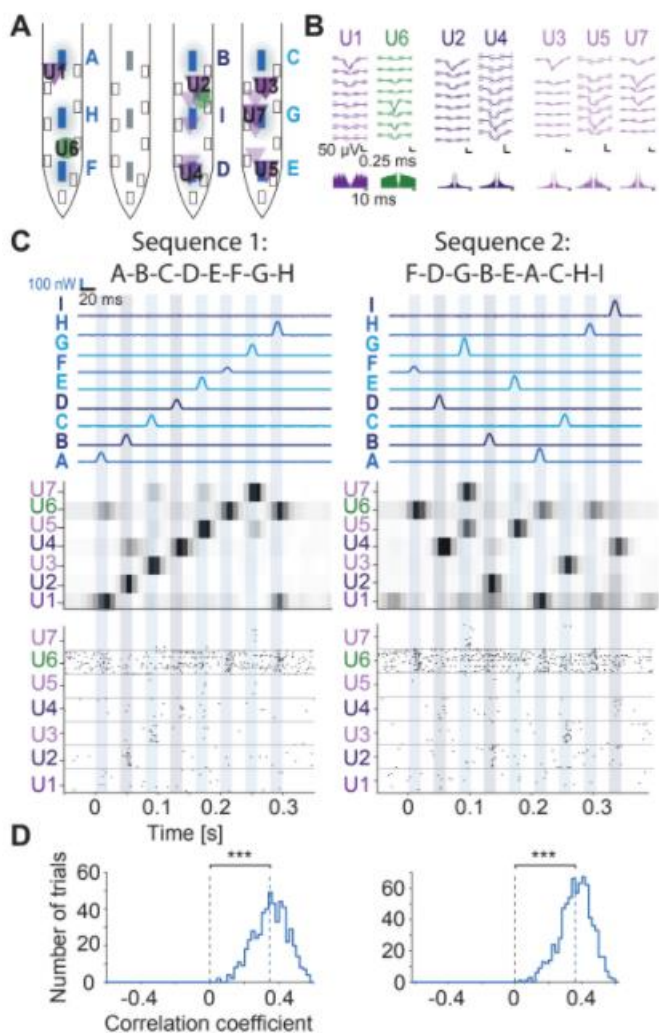

**Fig 9. Synthetic multi-neuronal sequences are generated using multi-site illumination.** (**A**) Schematic of implanted probe, same notation as **Fig. 7A**. Nine μLEDs (labelled A-H) and seven units (labelled U1-U7) were selected for sequence induction. (**B**) Wide-band spike waveforms and autocorrelation histograms of the selected simultaneously-recorded units. (**C**) Two different multi-μLED sequences were generated. *Top row*: illumination pattern. *Middle row*: PSTH for each of the units, averaged over 547 trials for sequence 1 (left) and over 750 trials for sequence 2 (right), and scaled to the 0-1 range. Every unit spiked with the highest gain during illumination by a specific μLED. *Bottom row*: raster plots of spikes during 30 trials. (**D**) Histograms show distributions of rank correlation coefficients between the mean illumination pattern and the single-trial firing rates. Vertical dashed lines indicate group medians (***: p<0.001, Wilcoxon signed rank test).

Localized illumination ranging between the spiking threshold and the population threshold allows controlling the activity of different units by distinct µLEDs. After mapping the spiking thresholds for units recorded next to every µLED, we chose µLEDs which induced spiking of individual units, and generated sequences of spikes in multiple units using temporal patterning of the µLEDs. During sequence induction, µLEDs that induced spiking activity were selected, and 15-20 ms light sinusoids were generated sequentially by the selected µLEDs, in a random order. At least two sequences were used within each session, for at least 500 trials. **Fig. 9** shows examples of such sequence control. In this case, nine µLEDs (three per shank, on three shanks) activated seven units in a sequence determined by the order of µLED activation (**Fig. 9C, left**). A random permutation of the activation sequence was followed by corresponding rearrangement of the multi-neuronal spiking sequence (**Fig. 9C, right**). To quantify sequence induction, rank correlation coefficients were calculated between the mean illumination pattern and the single-trial firing rates, obtained by convolving each spike with a Gaussian kernel ($\sigma$=5 ms). For both sequences, single-trial spike trains were consistently correlated with the illumination pattern. Overall, 15 sequence experiments were conducted during six sessions spaced over six weeks. The experiments involved a median of six units (range: 3-7). In all (15/15, 100%) experiments, consistent sequences were observed (p<0.05, Wilcoxon's signed-rank test, comparing to a zero median). The median rank correlation coefficient was 0.16 (n=15; p<0.001, Wilcoxon's signed-rank test). Thus, the system allows precise multi-neuronal spike patterns at the spatiotemporal resolution of individual neurons and millisecond timescale, in freely-moving mice.

## VI. DISCUSSION

We developed a 32 channel ASIC current source designed for sourcing up to 1 mA per channel against a 10 x 15 µm µLED load. We integrated the ASIC with all required peripherals into a headstage PCB we call "Outan", which can drive up to three 32-channel/12-µLED probes simultaneously. The system achieves linear control of the light power output of the µLEDs with <20 nW resolution by digitally calibrating every ASIC channel according to the properties of each µLED in a specific probe. A random access communication scheme allows serial control of 32 channels with 5 kHz refresh rate per channel, or higher with a smaller number of active channels. The system was tested with a freely-moving CaMKII::ChR2 mouse. Driving different µLEDs induced distinct LFP activity patterns, demonstrating the spatial resolution of the stimulation. Driving the same µLED at lower intensities induced spiking activity, while higher intensities induced population oscillations, demonstrating that the system has a wide dynamic range. Driving multiple µLEDs in arbitrary sequences induced spiking activity in multiple single units at corresponding sequences, demonstrating the high spatiotemporal resolution achievable by the system.

The presented system drives neuron-sized (10 x 15 µm) µLEDs, which provide a high spatial resolution, but have a higher forward voltage than the 200 µm sized µLEDs driven by other circuits ([20], [23] and [25]). While similar in design goals to [27], our ASIC differs in several aspects: (1) the DAC architecture (we used R-2R and an OTA, while a current steering DAC was used in [27]); (2) the handling of the large voltage drop on the transistors (we used voltage clamps controlled by the digital input, while [27] added diode-connected transistors); and (3) the digital addressing (we used random access, while serialized control of all the channels in every command was used in [27]). The maximum output current of our system was measured to be more than five times the results presented in [27]. In contrast to [24] and [27], our system has been demonstrated in experiments with a freely moving mouse, carried over multiple months. The proposed system is modular, allowing replacement of driving, recording or implantable modules for repair and upgrades. **Table II** summarizes the comparison between our and previous reports.

A notable limitation is that the driver system is completely feed-forward. Thus, the controller has no way of knowing how much light was actually delivered to the tissue. Although often indicative, the recorded neural signals cannot be used for reliable feedback since many other factors influence ongoing neuronal activity. Lack of precise feedback makes the system sensitive to glitches in the digital communication. When a communication error occurs, the wrong value persists until the mistakenly addressed channel receives another command. In addition, there is no feedback upon the integrity of a given µLED. A related limitation is that the system is designed specifically for driving µLEDs with a narrow range of I-V profiles. If a channel is activated while loaded with a broken shank (open) or with a burnt-out µLED (short), the effective load is outside the recommended range of voltages which may shorten the shelf-life of the ASIC.

TABLE II
COMPARISON TO STATE OF THE ART MINIATURE µLED DRIVERS

|  | [20] | [21]-[23] | [24]-[26] | [27] | **This work** |
|---|---|---|---|---|---|
| Technology | Discrete circuit | 130 nm | 350 nm | 180 nm | 65 nm |
| Number of channels | 4 | 4 | 18 | 48 | 32 |
| µLED size | 220 x 270 µm | 180 x 230 µm | 240 x 320 µm | 10 x 15 µm | 10 x 15 µm |
| µLED forward voltage | 2.9 V @ 20 mA | 3.36 V @ 20 mA | 3 V @ 3 mA | 5 V @ 1 mA | 5 V @ 1 mA |
| Max. measured current | 25 mA | 20 mA | 4.37 mA | 0.1 mA | 0.6-0.9 mA |
| Design resolution | 8 bits | 3 bits | 8 bits | 10 bits | 10 bits |
| Control scheme | Digital potentiometer | PWM | PWM | Parallel (480 bit code) | Random Access (16 bit code) |
| Probe integration | Omnetics connector | Doric connector | Same wafer | Same PCB | Omnetics connector |
| Animal demonstration | Freely moving rat | Freely moving mouse | Anesthetized mouse | Anesthetized mouse | Freely moving mouse |

In future versions of the ASIC, we propose adding an ADC that samples the output current and reports back to the controller. This will allow detection of damaged µLEDs and correction of communication errors. Second, implementing the voltage regulators and level shifters in the ASIC itself will not make the ASIC considerably larger, but may simplify the assembly of the Outan PCB and enable further miniaturization. Presently, the minimum size of the PCB is dictated by the connectors. At the system level, in addition to the recording, stimulation and probe PCBs, a processing PCB will be added. The processing PCB will read data from the recording PCB and command the stimulation PCB, creating a closed loop system that does not require tethers to a stationary system.

To conclude, the system is easy to use, allowing neuroscientists with little background in electronics to utilize the advantages of integrated µLED probes without building complicated custom back-end hardware. The in-vivo experiments presented in this paper demonstrate previously impossible interventions. The high spatial, temporal, and amplitude resolution of the system provides an unprecedented variety of illumination patterns. Activation of cells in a synthetic order provides an opportunity to derive causal conclusions about the function of neural circuits.

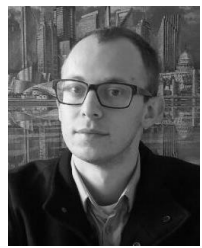

**Alexander Tarnavsky Eitan** Received the B.Sc. degree in physics and the B.Sc. degree in electrical engineering from the Technion–Israel Institute of Technology, Haifa, Israel, in 2011, and the M.Sc. degree (cum laude) in electrical engineering in 2020 from Tel Aviv University, Tel Aviv, Israel, where he is currently working toward a Ph.D. degree in Neuroscience.

From 2011 to 2018 he was a Research Officer with the Israel Defense Forces (IDF). His current research interests include systems neuroscience, neuro engineering and brain-machine interfaces.

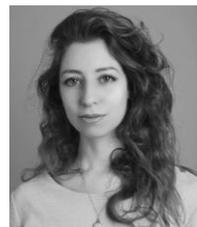

**Shirly Someck** received her B.Sc. in Physics and Philosophy from Tel Aviv University, Tel Aviv, Israel in 2016. Following that, she started the direct Ph.D. program in the Sagol School of Neuroscience at Tel Aviv University, Tel Aviv, Israel.

Her current research focuses on finding a necessary and sufficient neuronal pattern which induces memory-guided behavior, by utilizing extracellular recording, optogenetic manipulations and a novel memory task.

Ms. Someck is a Minducate fellow in the Sagol School of Neuroscience and Online Innovating Education center of Tel Aviv University, and a student member of the ISFN.

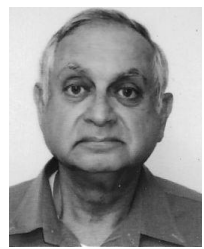

**Mario Guillermo Zajac** finished studies as an Electronics Technician in Telecommunication in 1971 from the Ort School in Buenos Aires, Argentina. Received the B.Sc. degree in electrical engineering from the Technion – Israel Institute of Technology, Haifa, Israel in 1984. Worked in R&D with Elscint (computerized tomography), Fibronics (fiber optics transmitters and receivers), Shorashim Medical (Electro Encephalography instrumentation), Petrometrix (Fuel Analysis and Spectroscopy), SCD (infrared video cameras interfacing), Intel (CMOS ADSL transmitter and WiFi baseband circuits design). From 2014 he is with Tel Aviv University. His research interests are with high frequency active filter design and high frequency operational amplifier design.

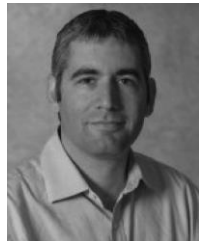

**Eran Socher** (Senior Member, IEEE) received the B.A. degree (summa cum laude) in physics and the B.Sc. (summa cum laude), M.Sc., and Ph.D. degrees in electrical engineering from the Technion–Israel Institute of Technology, Haifa, Israel, in 1996, 1996, 1999, 2005, respectively, with a focus on CMOS compatible MEMS sensors and actuators and their readout electronics, especially for uncooled thermal imaging. From 2003 to 2006, he was a Research Engineer with the Israel Defense Forces (IDF) and an Adjunct Lecturer with Technion and Bar-Ilan University, Ramat Gan, Israel. From 2006 to 2008, he was a Visiting Researcher with the High Speed Electronics Laboratory and a Visiting Assistant Professor with the Department of Electrical Engineering, University of California at Los Angeles (UCLA), Los Angeles, CA, USA. Since 2008, he has been with the School of Electrical Engineering, Tel Aviv University (TAU), Tel Aviv, Israel. He heads the High Frequency Integrated Circuits Laboratory, where he is currently an Associate Professor. He was a Visiting Professor with the Department of Electrical and Computer Engineering, University of Toronto, Toronto, ON, Canada, from 2015 to 2017. He has coauthored over 100 journal and conference papers. His research interests are currently focused on RF and millimeter-wave CMOS circuit design for high-data-rate communication, sensing, and imaging. Prof. Socher was a recipient of several teaching and research awards and scholarships, including the TAU Rector's Prize for Excellent Teaching (twice). He has served as an Associate Editor for the IEEE MICROWAVE AND WIRELESS COMPONENTS LETTERS since 2015.

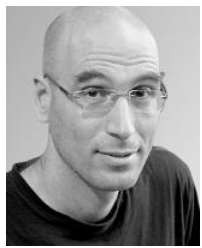

**Eran Stark** received the B.Med.Sc. and M.D. degrees in medicine and the Ph.D. (summa cum laude) degree in neural computation, in 1999, 2006, and 2008 respectively, all from the Hebrew University in Jerusalem, Israel. From 2008 to 2011 he was a Rothschild, Machiah, and HFSP scholar and a post-doctoral Research Fellow in Rutgers University, Newark, NJ (USA), with a focus on neuro engineering. From 2012 to 2014 he was a Research Assistant Professor in New York University, New York City, NY (USA), focusing on neurophysiology. Since 2015 he is with the department of Physiology and Pharmacology, Sackler School of Medicine, in Tel Aviv University, Tel Aviv, Israel, where he directs the Center of Excellence for the Study of the Neural Code underlying Cognition. His research interests include information transmission and processing at the spatiotemporal resolution of a single neuron and a single spike: within the nervous system, and between the nervous system and the external world